\documentclass[aps, prl, twocolumn, amsmath, amssymb,superscriptaddress]{revtex4-1}

\usepackage{graphicx}% Include figure files
\usepackage{dcolumn}% Align table columns on decimal point
\usepackage{bm}% bold math
\usepackage{mathrsfs}
\usepackage{amsmath}
\usepackage{color}
\DeclareMathOperator{\sign}{sign}
\begin{document}

%Title of paper
\title{Coherent excited states in superconductors due to a  microwave field}
\author{A.\,V.~Semenov}
\email[]{av.semyonov@mpgu.edu}
%\email[]{Your e-mail address}
%\homepage[]{Your web page}
%\thanks{}
%\altaffiliation{}
\affiliation{Moscow State Pedagogical University, 1 Malaya Pirogovskaya st., Moscow 119992, Russia}
\affiliation{Moscow Institute of Physics and Technology, Dolgoprudny, Moscow 141700, Russia}

\author{ I.\,A.~Devyatov}
\email[]{igor-devyatov@yandex.ru}
%\homepage[]{Your web page}
%\thanks{}
%\altaffiliation{}
\affiliation{Lomonosov Moscow State University,  Skobeltsyn Institute of Nuclear Physics, 1(2),  Leninskie gory, GSP-1, Moscow 119991, Russia}
\affiliation{Moscow Institute of Physics and Technology, Dolgoprudny, Moscow 141700, Russia}

\author{P. J. de Visser}
\affiliation{Kavli Institute of NanoScience, Faculty of Applied Sciences, Delft
University of Technology, Lorentzweg 1, 2628 CJ Delft, The Netherlands}
\affiliation{SRON Netherlands Institute for Space Research, Sorbonnelaan 2, 3584 CA Utrecht, The Netherlands}

\author{T. M. Klapwijk}
\email[]{T.M.Klapwijk@tudelft.nl}
\affiliation{Kavli Institute of NanoScience, Faculty of Applied Sciences, Delft
University of Technology, Lorentzweg 1, 2628 CJ Delft, The Netherlands}
\affiliation{Moscow State Pedagogical University, 1 Malaya Pirogovskaya st., Moscow 119992, Russia}

\date{\today}

\begin{abstract}

We describe theoretically the depairing effect of a microwave field on diffusive s-wave superconductors. The  ground state of the superconductor is altered qualitatively  in analogy to the depairing due to a dc current.  In contrast to dc-depairing the density of states acquires, for microwaves with frequency $\omega_0$, steps at multiples of the photon energy $\Delta\pm n\hbar\omega_0$ and shows an exponential-like tail in the subgap regime. We show that this ac-depairing explains the measured frequency shift of a superconducting resonator with microwave power at low temperatures.
\end{abstract}

\maketitle

How is the superconducting state modified by a current, \emph{i.e.} when the condensate is moving?
The answer to this question is well-known for the case of  a dc current flowing in a superconducting wire.
For a dc current, the Cooper pairs gain a finite momentum which leads to the suppression of the superconducting properties of the wire \cite{Tinkham,ant}.
The modulus of the superconducting order parameter $\Delta$ is reduced and the sharp BCS singularity near the gap is smeared. This depairing effect of a current or of a magnetic field was studied theoretically soon after the creation of the microscopic theory of superconductivity \cite{bar}.
The moving superconducting condensate has been called a coherent excited state generated by the momentum displacement operator $\rho_q=\sum_n \exp(i \textbf{q} \cdot \textbf{r}_n)$ by Anderson \cite{andr} as part of the explanation of the Meissner effect from the original form of the BCS theory.
The momentum displacement operator, when applied to the BCS ground state, creates excited pairs of electrons $\textbf{k}_1$, $\textbf{k}_2$ with momentum pairing $\textbf{k}_1 + \textbf{k}_2 = \textbf{q}$ instead of zero \cite{andr,DeGennes, Tinkham, bag}. This momentum displacement $q = |\textbf{q}|$ corresponds to a superfluid drift velocity $v_s=\hbar q/m$, where $m$ is the electron mass.
In the Green's function technique it is possible to introduce the superfluid velocity in a gauge invariant way $v_s \propto [\nabla \varphi - (2e/\hbar)\textbf{A}]$, where $\varphi$ is the phase of the superconductor, $e$ the electron charge and $\textbf{A}$ the vector potential of the electromagnetic field. The equivalence of depairing due to an electric current and due to a magnetic field is well established, both theoretically\cite{maki} and experimentally\cite{ant}, using thin and narrow superconducting wires with a uniform current-density. The theory of depairing by a dc current was reformulated, using the Usadel equations \cite{us}, for diffusive films with an elastic scattering length much smaller than the BCS coherence length \cite{kupr1}. The results of this theory \cite{kupr1} were confirmed experimentally by Romijn et al\cite{rom} and by Anthore et al\cite{ant}.

However, a general theory for \emph{depairing by a microwave field}, a time-dependent vector potential $\textbf{{A}}$, has not been formulated. In current experimental research there are many cases in which a superconductor is used at very low temperatures, $T/{T_c} \ll 1$, where the density of quasiparticles is very low and the response of the superconductor is dominated by the response of the superfluid. At higher temperatures it is well known that microwave radiation can be absorbed by quasi-particles, leading to a non-equilibrium distribution over the energies\cite{elias1972}. At very low temperatures there are hardly any quasiparticles and with $\hbar\omega_0 \ll 2\Delta$ there is not enough energy per photon to break Cooper-pairs.  In practice it has been demonstrated that in this regime the microwave power has a significant influence on the superconducting state. For  example De Visser et al\cite{dv,dv2} have measured for aluminium a shift in kinetic inductance and an increase in the density of quasiparticles as a function of microwave power at temperatures of 60 mK. This dependence can be parametrized by labeling it as 'absorbed power', but the fundamental question is how the superconducting state responds to a time-dependent electromagnetic field. In previous works only the limiting case of high temperatures (close to the critical temperature $T_c$) \cite{kul} or relatively low frequency compared to the temperature ($\hbar\omega_0 \ll k_B T$) \cite{gur} were studied. In these cases the coherent properties of a superconductor (e.g. density of states) change analogous to magnetic impurities and a static magnetic field. However, in general it is to be expected that an oscillating vector-potential would lead to coherent excited pairs with an oscillating center of mass motion, with a more substantial modification of the density of states. The change of the energy spectrum of electrons, dressed by an electromagnetic field, is in principle analogous to the dynamic Stark effect in atom physics\cite{stark} (and similarly for a two-dimensional electron gas\cite{morina}).  Such dressed states have recently been put forward in the analysis of the microwave response of superconducting atomic point contacts\cite{ber1,ber2}. The relevance of the problem for a plain superconductor is apparent in the case of superconducting parametric amplifiers \cite{eom} and in the nonlinearity of kinetic inductance devices with microwave-readout \cite{day,zm,dv2}. Additionally, the effects of direct depairing at these frequencies is also of interest for quantum optics on superconducting artificial quantum systems \cite{astafiev2010,astafiev2012}. In summary, there is an urgent need for a microscopic theory which correctly describes the depairing of a conventional superconductor by microwave radiation.

We focus on the influence of high-frequency radiation on a narrow and thin dirty s-wave superconducting strip  using the well-established theory of nonequilibrium superconductivity \cite{laov,bel}. Diffusive motion implies that the direction of momentum is randomized on the period of the microwave field. The thickness and width are assumed to be much less than the relevant penetration depth $\lambda$, allowing for a uniform current over the cross-section. The strip is assumed to be long enough in order to ignore the influence of the ends of the strip. We consider the case of a relatively high frequency $\omega_0$  of the microwave radiation:
\begin{equation}
\alpha \ll \hbar\omega_0 \ll 2\Delta,
\label{lim}
\end{equation}
where the parameter $\alpha$ is the same as in the Eliashberg  theory \cite{elias1972}: $\alpha=e^2DE^2_0/\hbar\omega^2_0$, with $D$ the diffusion coefficient and $E_0$ the amplitude of the microwave field.
The left inequality of Eq. (\ref{lim}) is known as the condition for the quantum regime of absorption \cite{elias1972,sh,tie,tuck}, and by analogy we call this regime of depairing the 'quantum regime of depairing'. The right inequality of Eq. (\ref{lim}) points out that it is not possible to excite quasiparticles directly. Hence, the depairing is only associated with the reconstruction of the ground state of the superconductor in the presence of a microwave field.

We start from the usual mean-field Hamiltonian of a superconductor in the presence of electromagnetic radiation, modeled by a vector-potential $\textbf{A}$, introduced in the gauge-invariant form in the kinetic part of this Hamiltonian:
$H_k=-\frac{\hbar^2}{2m}\partial^2,~\partial=\partial/\partial \textbf{r}-ie\textbf{A}\check{\tau}_z$
where $\check{\tau}_z$ is the third Pauli matrix, operated in Keldysh-Nambu spaces \cite{laov,bel}.
We use the Keldysh-Usadel approach \cite{laov,bel,sem1} to determine the response of the superconductor. For details we refer to the Supplemental Material\cite{som}. Under the condition of Eq. (\ref{lim}), the spatially homogeneous dirty s-wave superconductor can be described by a closed set of equations containing only the stationary components of the Green's functions and the order parameter. The stationary components of the retarded Green's functions satisfy in the presence of a dc current and in the presence of monochromatic radiation the equation:
\begin{equation}
-iEF^R_0-i\Delta G^R_0 +\Pi =0,\\
\label{cf1}
\end{equation}
In Eq.~(\ref{cf1}),  $G^R_0=G^R_0(E)$ and $ ~F^R_0=F^R_0(E)$ are energy ($E$) dependent stationary normal and anomalous components of the superconducting matrix Green's function $\hat{G}^R(E)$ in Nambu space. In  the case of  depairing by a dc current\cite{ant}, the depairing term $\Pi$ in Eq. (\ref{cf1}) has the form
$\Pi=\Gamma G^R_{0}F^R_{0}$ , where $\Gamma$ is the depairing energy determined by the current-density.
In the present case of monochromatic radiation, modeled by a vector-potential $A(t)=A\cos(\omega_0t)$ the depairing term $\Pi$ is given by:
\begin{equation}
\Pi=\alpha\{F^R_0(G^R_{0+}+G^R_{0-})+G^R_0(F^R_{0+}+F^R_{0-}) \}.
\label{crucialterm}
\end{equation}
Equation~\ref{crucialterm} constitutes the formal difference of depairing by an rf current from the case of depairing by a dc current.  $G^R_{0\pm}=G^R_0(E\pm \hbar\omega_0)$ and $F^R_{0\pm}=F^R_0(E\pm \hbar\omega_0)$, are both shifted in energy by $\hbar\omega_0$ with respect to the normal and anomalous component of the Green's function. $G^R_0$ and $F^R_0$ are linked via the normalization condition
\begin{equation}
(G^R_0)^2-(F^R_0)^2=1.
\label{cf2}
\end{equation}

 \begin{figure}
 \includegraphics[width=0.99\columnwidth]{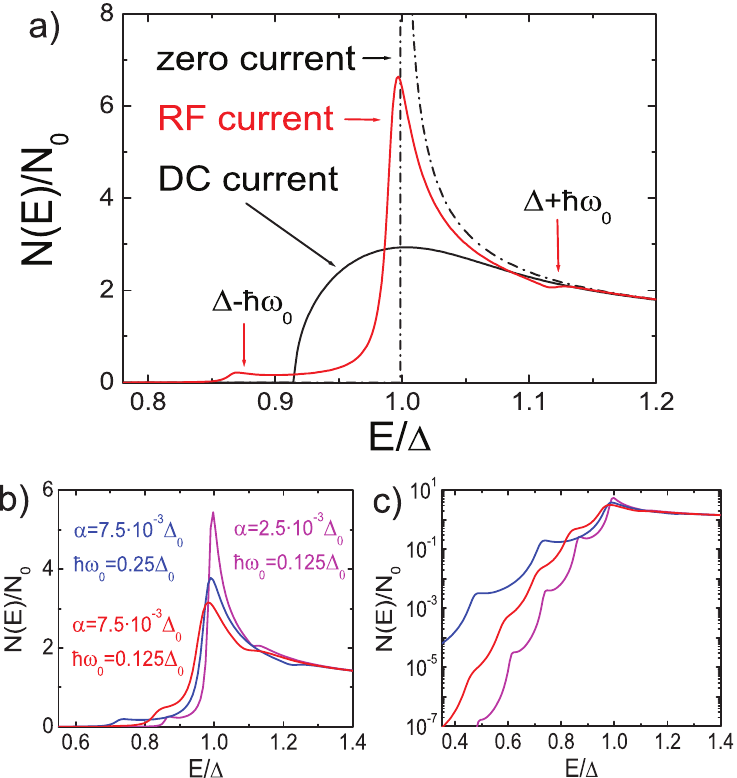}
 \caption{The normalized density of states of a diffuse superconducting strip $N(E)/N_0=\Re G_0^R(E)$, subject to a monochromatic microwave signal. The black dotted line in Panel a corresponds to the pure BCS  case without any depairing $\alpha=\Gamma=0$. The black solid line corresponds to depairing by a dc current with the value of the depairing parameter $\Gamma=0.014 \Delta_0$. The red curve in Panel a corresponds to rf depairing with the value of the parameter $\alpha=0.0017\Delta_0$. The red, blue and magenta curves in the panels b) and c) correspond to rf depairing with different values of and for different frequencies: the red curve corresponds to $\alpha=0.0075 \Delta_0,~ \hbar\omega_0=0.125 \Delta_0$, the
blue curve corresponds to $\alpha=0.0075 \Delta_0,~ \hbar\omega_0=0.25 \Delta_0$, and the
magenta curve corresponds to $\alpha=0.0025 \Delta_0,~ \hbar\omega_0=0.125 \Delta_0$. Panel c shows on a logarithmic scale the presence of the low-lying states, which are absent for zero or DC current, and steps at multiples of $\hbar\omega_0$.}
    \label{dens}
\end{figure}

The order parameter $\Delta$, also stationary, satisfies the self-consistency equation of the usual form
\begin{equation}
\Delta=-\lambda_{ep}\int\limits_{0}^{\hbar\omega_{D}}dE\left(
1-2f_0\right)  \Re F^{R}_0,\label{delta}
\end{equation}
and is thus field-dependent, where $f_0=f_0(E)$ is the stationary component of the distribution function, $\omega_{D}$ the Debye frequency, and $\lambda_{ep}$ the electron-phonon coupling constant. In the case of excitations by microwaves the distribution function is symmetric in energy, $f(E)=f(-E)$, i.e. only the longitudinal part $f_L$ of the distribution function arises: $\sign(E)(1-2f(E))=f_L(E)$\cite{laov,bel}.

This set of equations (Eqs. (\ref{cf1}) - (\ref{delta})) has been solved numerically, the procedure is described in the Supplemental Material\cite{som}. Fig. \ref{dens} shows the results for the density of states,  $N(E)=N_0\Re G_0^R(E)$, for different microwave intensities and frequencies (here $N_0$ is the density of states in the normal metal per spin). The black dashed-dotted line in Fig. \ref{dens}a corresponds to the pure BCS  case without any depairing, $\alpha=\Gamma=0$. The black solid line corresponds to depairing by a dc current with the value of the depairing parameter \textbf{$\Gamma=0.014 \Delta_0$}. $\Delta_0$ is the modulus of the order parameter at zero temperature without depairing ($\alpha=0$). The red curve corresponds to depairing by microwaves for  $\alpha=0.0017 \Delta_0$, with the photon-energies clearly visible. To facilitate a quantitative estimate we express the ratio $\Gamma/\Delta_0$ as the ratio of the dc supercurrent $j$ to the critical pair-breaking current $j_c$ with $\Gamma/\Delta_0 \cong 0.11(j/j_c)^2$. Similarly, for $\alpha$: $\alpha/\Delta_0 \cong 0.014(j_0/j_c)^2$, with $j_0$ the amplitude of the rf supercurrent. The DC current and RF current curves in  Fig. \ref{dens}a correspond to the same value of $j$, $j_0=0.25 j_c$. Fig. \ref{dens}b presents a comparison between densities of states for different values of the depairing parameter $\alpha$ and different frequencies, as indicated in the figure.

As is evident from Fig. \ref{dens} the depairing due to the rf signal (red, blue and magenta curves) is very different from the depairing due to a dc current (black curve) although both smear the BCS singularity at $E=\Delta_0$. One striking difference is that the density of states, which corresponds to dc depairing, goes to zero at some value of energy, whereas the densities of states, which corresponds to rf depairing have exponential-like tails at small energies (shown in Fig. \ref{dens}c).
In addition, with rf depairing there are irregularities in the density of states at equal distance from each other, corresponding to the photon energy $\hbar \omega_0$ (Fig. \ref{dens}). These photon-steps are reminiscent  of the photon-assisted tunneling steps in the quasi-particle current of tunnel junctions irradiated by a microwave signal \cite{tie,tuck}. In the present case for a plain superconducting film and frequencies corresponding to the right inequality in Eq. (\ref{lim}), the photon-structures in the density
of states can be interpreted as a manifestation of dressed states of diffusely scattering electrons, which inevitably must exist in a superconductor in a microwave field. This would naturally arise in the displacement operator $\rho_q$ by replacing a time-independent $q$ by $q_0 \cos \omega t$. Here, this is accomplished for a diffusive superconductor leading to Eqs.~\ref{cf1} and \ref{crucialterm}.  

Qualitatively, the result reflects that the dressed electron states become superpositions of states with energies shifted by multiples of $\hbar\omega_0$. The transition probability of a diffusely moving electron from the state with energy $E$ to the state with energy $E \pm \hbar\omega_0$ per unit time is of order $\alpha/\hbar$ \cite{elias1972}, thus during the oscillation period of the field, $2\pi/ \omega_0$, the components with energy shifted to $\pm \hbar\omega_0$ acquire a weight of order $\alpha/\hbar\omega_0$. Transitions to $\pm n \hbar \omega_0$ are processes of $n$-th order and the corresponding weights are of order $(\alpha/\hbar\omega_0)^n$. From this consideration, it follows that the BCS peak in the density of states resurfaces in the series of peaks at the energies of photon $\Delta \pm n\hbar\omega_0$ (photon points), of which the amplitudes decrease exponentially with the growth of $n$.

\begin{figure}[ht]
\includegraphics[width=0.99\columnwidth]{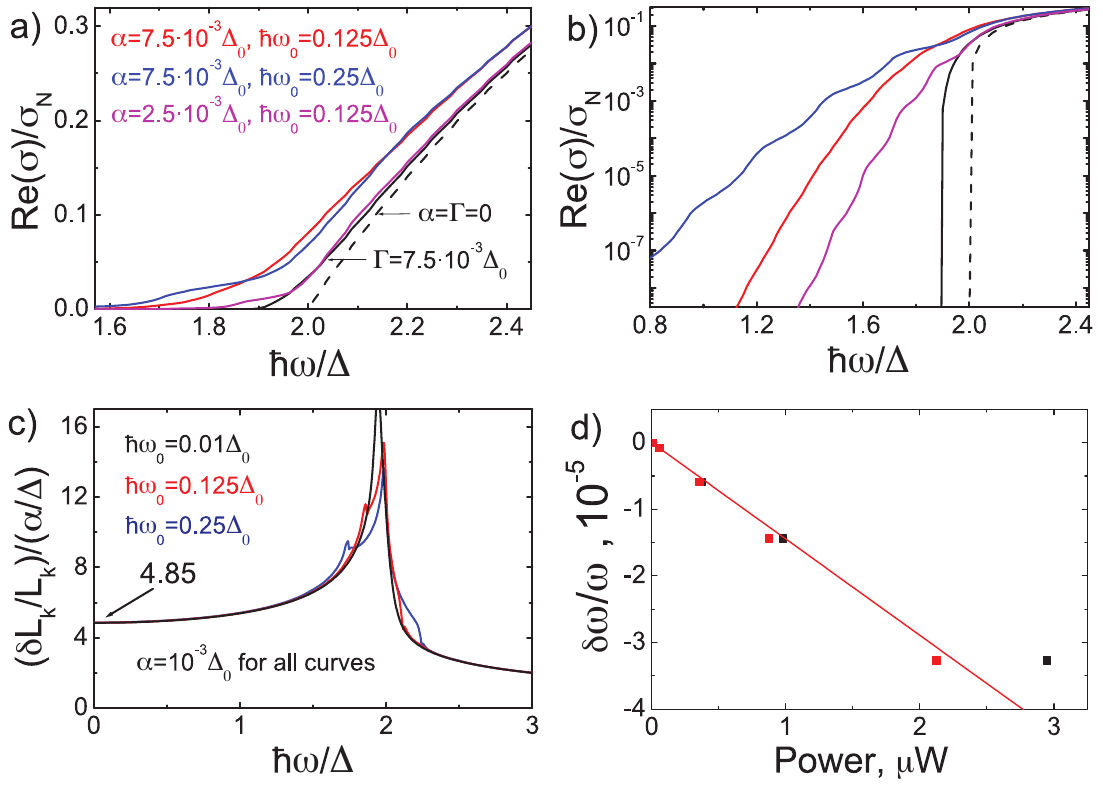}
\caption{(a) The real part of the conductivity $\Re \sigma(\omega)$  with the same values of depairing parameters $\alpha,\Gamma$ and signal frequencies $\omega_0$ as for the densities of states in  Fig. \ref{dens}b,c. (b) The real part of the conductivity on a logarithmic scale to highlight the effect of the low-lying states. The colour coding of panel (a) applies. (c) The changes in the imaginary part of the conductivity due to the microwave power for values of $\alpha$ and $\hbar\omega_0$ as indicated in the figure. (d)  The observed shift of the resonant frequency of a superconducting resonator as a function of the power of the microwave radiation. The red line shows the shift of the resonant frequency as a function of the internal power in the resonator, calculated using Eq. (\ref{delta2}). Black squares represent the experimental data based on data analysis with a simple Lorentzian resonance curve\cite{dv}. The red squares correspond to the experimental data \cite{dv}, based on an improved analysis by taking into account the nonlinearity of the kinetic inductance.}
\label{dvi}
\end{figure}

Apart from its fundamental interest this theoretical analysis is relevant for experiments on microwave kinetic inductance detectors and parametric amplifiers based on superconducting films at milliKelvin temperatures, where the electrodynamic response is dominated by the superconducting condensate. Typically, the quantity which is measured is the impedance or the complex conductivity. The conductivity $\sigma(\omega)$ at frequency $\omega$ can be derived from the general formula for the current density in a diffusive superconductor in a straightforward way\cite{som}:
\begin{equation}
\begin{aligned}
&\sigma(\omega)=\frac{\sigma_N}{4 \omega} \int d E \{
((G_{0-}^{R})^* \Re G_0+(F_{0-}^{R})^* \Re F_0)f_{L0}+\\
&(G_0^R \Re G_{0-}^R+F_0^R \Re F_{0-}^R)f_{L0-}\},\\
\end{aligned}
\label{sigma}
\end{equation}
with $\sigma_N=2 e^2 N_0 D$ the conductivity of the wire in the normal state and $f_{L0-}=f_{L0}(E-\hbar \omega_0)$. In deriving Eq. (\ref{sigma}) we ignore higher order terms leading to a modulation of the conductivity at the doubled signal frequency. Fig. \ref{dvi}a,b shows the real part of conductivity $\Re \sigma(\omega)$, calculated from Eq. (\ref{sigma})  with the same values of the depairing parameters $\alpha,\Gamma$ and signal frequencies $\omega_0$ as used in Fig. \ref{dens}b and c,  for the densities of states. In Fig. \ref{dvi}a,b one observes, qualitatively, the same behavior in the real part of the conductivity $\Re \sigma(\omega)$ as for the densities of states: the microwave radiation leads to the appearance of anomalies at 'photon points' and an exponential-like tail at small values of $\omega$.
In a recent experiment\cite{dv} a superconducting resonator of aluminium was measured to determine the quality factor and the resonant frequency as a function of applied microwave power. In the interpretation the absorption by quasi-particles was included, through the well-known approach introduced by Eliashberg\cite{elias1972}. However, it was found that, at the lowest temperatures, it is not possible to account for the shift in resonant frequency.  The observed shift was stronger than one would expect on the basis of the shift in quality-factor, which is dominated by the redistribution of quasi-particles over the energies.  In the present theoretical analysis the new ingredient is the modification of the superconducting ground state in the presence of microwaves. Therefore we focus on the change of resonant frequency with microwave power, which is a measure of the change in kinetic inductance, \emph{i.e.} the Cooper-pair density, whereas the quality factor is a measure of the losses due to the quasi-particles. The kinetic inductance is given by the imaginary part of the conductance, which is shown in Fig. \ref{dvi}c for a selection of $\alpha$ and $\omega_0$. Substitution of the calculated Green's functions into the imaginary part of Eq. (\ref{sigma}) gives, for low values of $\alpha$, low frequencies
(corresponding to the condition (\ref{lim})) and low temperatures $T \ll T_c$ a simple linear relation between $\alpha$ and the shift of kinetic inductance $\delta L_k$ with respect to its value without rf field $L_k$:

\begin{equation}
\begin{aligned}
\delta L_k/L_k \cong 4.85 \alpha/\Delta_0.
\end{aligned}
\label{delta2}
\end{equation}
To facilitate the comparison of Eq. (\ref{delta2}) with the recent experiments \cite{dv}, we rewrite Eq. (\ref{delta2}) in the quantities which were measured in the experiment: the relative shift of the resonant frequency $\delta \omega/\omega \cong - 2.42 P_{in}/P_0$. Here, $P_{in}$ is the (internal) power of the microwave radiation in the resonator and $P_0=2 \pi Z_0 N_0 \Delta_0^3 w^2 d^2/(\hbar \rho)$ is the material parameter which relates the critical current to power, with $Z_0$ the impedance of the coplanar waveguide, $\rho$ the normal state resistivity of the superconducting resonator (Al), and $w$ and $d$ the width and thickness of the superconducting strip, respectively. This conversion was done under the assumption of a uniform distribution of the supercurrent across the strip.

In Fig. \ref{dvi}d we show the shift of the resonant frequency as a function of microwave power calculated using Eq. (\ref{delta2}) as the red line. In the same figure the black squares correspond to the experimental data of De Visser et al\cite{dv}, determined for a resonant frequency of 5.3 GHz, well above $kT$ (64 mK) and well below $\Delta$ (2.2 K). We include the experimental data as black squares, which agree very well with the theoretical prediction, except at the highest power. However, in the analysis of the measured resonance curves the non-linearity in the kinetic inductance was not taken into account. By including  the nonlinearity of the kinetic inductance in the analysis of the measured resonance curves following Swenson et al\cite{swen} the black data-points from \cite{dv} were replaced by the red data points (Fig. \ref{dvi}d), which agrees very well with the theoretical prediction of Eq. (\ref{delta2}) without using fitting parameters.

These results have also implications for the density of thermal quasiparticles for a given temperature.  Obviously the reconfiguration of the ground state in the presence of microwave power reduces the condensation energy and therefore we expect for a given bath temperature a rapid increase in the number of quasiparticles. The number of excess quasiparticles $N_{qp}$ can be measured by further detailed studies of their fluctuations in a superconducting strip\cite{dv2}. We note that these other observables of the experiment \cite{dv} (number of quasiparticles, recombination time, quality factor), which more dominantly depend on $f(E)$ as compared to $\delta \omega/\omega$, are still dominated by a redistribution of the quasiparticles due to microwave absorption (Eliashberg\cite{elias1972}). In contrast $\delta \omega/\omega$ is a measure of the Cooper-pair current.

A direct test of the present theory would be a measurement of the density of states by tunneling. In such an experiment the time-dependent electric field should only be in one of the electrodes parallel to the tunnel-junction and not across the tunnel-junction. This requirement is to avoid that the dc tunneling process in itself is controlled by an additional ac electric field across the tunnel barrier, the well-known effect of photon-assisted tunneling \cite{tie,tuck}. A dc electric field should be across the tunnel junction to probe the electrodes with the tunneling current. The device could in practice be made as a transmission-line. A solution for such an experiment has been implemented by Horstman and Wolter\cite{hw1,wh1}, which is very feasible with the current level of technology. Instead of the nonequilibrium distribution function one should extract the density of states, similarly to the experiment for dc pairing carried out by Anthore et al\cite{ant}.

In summary, we have presented a theoretical analysis of the quantum depairing effect by a microwave field for a long thin strip of a diffusive s-wave superconductor. We have demonstrated that the density of states loses a sharp peak at the gap energy and acquires features at photon points $\Delta \pm n\hbar\omega_0$. Also  we have shown that the density of states develops an exponential-like tail in the sub-gap region. Both phenomena are in strong contrast to the case of depairing by a dc current. We have demonstrated that the predicted effect is responsible for the shift of the resonant frequency of an Al superconducting resonator by microwave power\cite{dv}.

\begin{acknowledgments}
We acknowledge financial  support from the Ministry of Education and Science of the Russian Federation, contract N 14.B25.31.0007 of 26 June 2013. Work of AS was performed within the state task project no. 2575. AS also acknowledges support from Russian Foundation for Basic Researches, grant no. 16-29-11779 ofi-m. TMK acknowledges the financial support from the European Research Council Advanced grant no. 339306 (METIQUM).
\end{acknowledgments}

\clearpage
\onecolumngrid
\setcounter{figure}{0}

\renewcommand{\thefigure}{S\arabic{figure}}

\section*{Supplemental material for: Coherent excited states in superconductors created by microwave field}

\subsection*{A.V. Semenov, I.A. Devyatov, P.J. de Visser and T.M. Klapwijk} 
%\date{\today}

\maketitle

Here we present a detailed derivation of the equations presented in the main text.\\

%The problem of describing kinetic-inductance detector receiving small signal and red out by large rf bias can be splitted into two parts: 1) calculation of distribution function, spectral quantities and impedance of the absorber in a background nonequilibrium state formed by readout signal; and 2) calculation of small corrections to the above mentioned quantities due to the signal.

\textbf{\S 1. Primary quantities and equations.}
All the information about the superconducting state in the Keldysh formalism is contained in a quasiclassical isotropic Green's function $\breve{G}$ which is a matrix both in Keldysh and Nambu spaces. Its structure in Keldysh space is
\[\breve{G}=\left(\begin{array}{cc} {\hat{G}^{R} } & {\hat{G}^{K} } \\ {0} & {\hat{G}^{A} } \end{array}\right). \]
The components $\hat{G}^{R,A,K} $ are matrices in Nambu space and linked to each other by relations that follow from their analytic properties
%\[\hat{G}^{R} =\left(\begin{array}{cc} {G^{R} } & {F^{R} } \\ {-F^{R} } & {-G^{R} } \end{array}\right),\]
\[\hat{G}^{A} =-\hat{\tau }_{3} \left(\hat{G}^{R} \right)^{+} \hat{\tau }_{3} .\]

The matrix Green's function obeys the following equation of motion (the Usadel equation in the Keldysh technique, derived by Larkin and Ovchinnikov \cite{laov}):
\begin{equation} \label{BasicUsadel}
\begin{array}{c}
{e^{2} D\left\{ A\breve{\tau }_{3} \circ \breve{G}\circ A\breve{\tau }_{3} \circ \breve{G}-\breve{G}\circ A\breve{\tau }_{3} \circ \breve{G}\circ A\breve{\tau }_{3} \right\}+}
 \\
{+\breve{\tau }_{3} \partial _{t_{1}} \breve{G}+\partial _{t_{2}} \breve{G}\breve{\tau }_{3} -i\left[\breve{\Delta}\circ ,\breve{G}\right]_{-} =-i\left[\breve{\Sigma }_{inel} \circ ,\breve{G}\right]_{-} }.
\end{array}
\end{equation}
The dependence on spatial coordinates, if any, is assumed to be removed due to an appropriate gauge transformation. Notations are standard, convolutions are
\[\left(a\circ b\right)\left(t_{1} ,t_{2} \right)=\int dt'a\left(t_{1} ,t'\right)b\left(t',t_{2} \right) ,\]
\[\left(A\circ G\right)\left(t_{1} ,t_{2} \right)=A\left(t_{1} \right)G\left(t_{1} ,t_{2} \right).\]
Terms containing $A$, $\breve \Delta$ and $\breve \Sigma_{inel}$ describe, correspondingly, the action of the electromagnetic field, superconducting pairing and scattering due to inelastic interactions.

Eq. \eqref{BasicUsadel} contains two equations with a different physical meaning: an equation for the retarded Green's function $\hat{G}^{R} $ (and the hermitian conjugated equation for the advanced Green's function $\hat{G}^{A} $) which allows to find spectral quantities, and a kinetic equation which allows to find the distribution function of quasiparticles.

%It can also be demonstrated that
Besides the Usadel equation the Green's functions obey the %following
normalization condition:

\begin{equation} \label{BasicNormCond}
\breve{G}\circ \breve{G}=\breve{1}\delta \left(t_{1} -t_{2} \right).
\end{equation}

%\textit{ \textbf{Note.} Counting number of scalar equations contained in \eqref{BasicUsadel} and \eqref{BasicNormCond}, one finds that it is twice as large as a number of components in $\breve{G}$. This means that some equations for components are identical to each other or identical to zero, as we will see this below.}

%\textbf{\S 2. Equation to find spectral quantities.} First consider
The retarded part of the Usadel equation \eqref{BasicUsadel}
%because it has simpler structure and in some sense closed with respect to $\hat{G}^{R} $, so it should be solved first. It
reads
\begin{equation} \label{R-Usadel}
\begin{array}{c}
{e^{2} D\left\{A\hat{\tau }_{3} \circ \hat{G}^{R} \circ A\hat{\tau }_{3} \circ \hat{G}^{R} \right. -\hat{G}^{R} \circ A\hat{\tau }_{3} \circ \hat{G}^{R} \circ \left. A\hat{\tau }_{3} \right\}+}
%\frac{\Gamma}{2}\left\{\hat{\tau }_{3} \hat{G}^{R}\circ \hat{\tau }_{3} \hat{G}^{R}-\hat{G}^{R}\hat{\tau }_{3}\circ\hat{G}^{R}\hat{\tau }_{3}\right\}+}
 \\
{+\hat{\tau }_{3} \partial _{t_{1} } \hat{G}^{R} +\partial _{t_{2} } \hat{G}^{R} \hat{\tau }_{3} -i\left[\hat{\Delta }\circ ,\hat{G}^{R} \right]_{-} =0}. \end{array}
\end{equation}
One can see that it does not contain the distribution function explicitly. The latter affects the solution of \eqref{R-Usadel} only via order parameter $\hat\Delta$ which depends on the distribution function and on $\hat{G}^{R}$ via the self-consistency equation. Corresponding normalization condition is $\hat{G}^{R}\circ \hat{G}^{R}=\hat{1}\delta \left(t_{1} -t_{2} \right)$. In the following evaluations, superscript $R$ will be omitted.\\

\textbf{Rewriting the Usadel equation for discrete real frequencies.}
%\textit{(there generally are mathematical evaluations, the only little physical idea appears in the last paragraph).}

For evaluation, it is useful to transit from the two-temporal representation for Green's functions to the two-energy representation, which is introduced via Fourier transformation
\[a\left(E,E-\omega \right)=\int dt_{1} dt_{2} \exp \left(iEt_{1} -i\left(E-\omega \right)t_{2} \right) a\left(t_{1} ,t_{2} \right).\]

In this representation, convolutions read:
\[\left(a\circ b\right)\left(E,E-\omega \right)=\int \frac{d\omega '}{2\pi } a\left(E,E-\omega '\right)b\left(E-\omega ',E-\omega \right) , \]
\[\left(A\circ G\right)\left(E,E-\omega \right)=\int \frac{d\omega '}{2\pi } A\left(\omega '\right)G\left(E-\omega ',E-\omega \right) ,\]
\[\left(G\circ A\right)\left(E,E-\omega \right)=\int \frac{d\omega '}{2\pi } G\left(E,E-\omega '\right) A\left(\omega -\omega '\right)\equiv \int \frac{d\omega '}{2\pi } G\left(E,E-\omega +\omega '\right) A\left(\omega '\right),\]
\[\partial _{t_{1} } \to -iE,\partial _{t_{2} } \to i\left(E-\omega \right).\]

Here $A(\omega)$ is the Fourier transform of $A(t)$,
\[A\left(\omega \right)=\int dt\exp \left(i\omega t\right)A\left(t\right) .\]

First we consider terms in the retarded Usadel equation that contain $A$. For compactness, some coefficients will be omitted while evaluating and restored later, when required. Substituting the expressions for convolutions and treating them as $\left(A\circ \hat{G}\circ A\circ \hat{G}\right)=\left(A\circ \hat{G}\right)\circ \left(A\circ \hat{G}\right)$ etc., one can see that they are
\[\left(A\circ \hat{G}\circ A\circ \hat{G}\right)\left(E,E-\omega \right)=\int \frac{d\omega _{1} }{2\pi } \frac{d\omega _{2} }{2\pi } \frac{d\omega '}{2\pi } A\left(\omega _{1} \right)\hat{G}\left(E-\omega _{1} ,E-\omega '\right)A\left(\omega _{2} \right)\hat{G}\left(E-\omega '-\omega _{2} ,E-\omega \right), \]
\[\left(\hat{G}\circ A\circ \hat{G}\circ A\right)\left(E,E-\omega \right)=\int \frac{d\omega _{1} }{2\pi } \frac{d\omega _{2} }{2\pi } \frac{d\omega '}{2\pi } \hat{G}\left(E,E-\omega '+\omega _{1} \right)A\left(\omega _{1} \right)\hat{G}\left(E-\omega ',E-\omega +\omega _{2} \right)A\left(\omega _{2} \right). \]

To make the physical meaning more explicit, one can transit to frequency-energy representation using the rule
\[G\left(E,E-\omega \right)=G_{\omega } \left(E-\omega /2\right),\]
where the argument in brackets has a meaning of energy and the argument in subscript has a meaning of frequency, which is the Fourier-conjugate of time. (Note that the energy is a half-sum of ``front'' and ``rear'' energies and the frequency is their difference).

Shifting the energy argument by $+\omega /2$, one rewrites the above-written terms in frequency-energy representation as
\[\left(A\circ \hat{G}\circ A\circ \hat{G}\right)_{\omega } \left(E\right)=\int \frac{d\omega _{1} }{2\pi } \frac{d\omega _{2} }{2\pi } \frac{d\omega '}{2\pi } A\left(\omega _{1} \right)\hat{G}_{\omega '-\omega _{1} } \left(E-\frac{\omega _{1} }{2} -\frac{\omega '}{2} +\frac{\omega }{2} \right)A\left(\omega _{2} \right)\hat{G}_{\omega -\omega '-\omega _{2} } \left(E-\frac{\omega _{2} }{2} -\frac{\omega '}{2} \right)\] \[\left(\hat{G}\circ A\circ \hat{G}\circ A\right)_{\omega } \left(E\right)=\int \frac{d\omega _{1} }{2\pi } \frac{d\omega _{2} }{2\pi } \frac{d\omega '}{2\pi } \hat{G}_{\omega '-\omega _{1} } \left(E+\frac{\omega _{1} }{2} -\frac{\omega '}{2} +\frac{\omega }{2} \right)A\left(\omega _{1} \right)\hat{G}_{\omega -\omega '-\omega _{2} } \left(E+\frac{\omega _{2} }{2} -\frac{\omega '}{2} \right)A\left(\omega _{2} \right). \] \\

\pagebreak
\textbf{The case of monochromatic field.}

Till the moment, the evaluations were simply rewriting of initial expressions from the retarded Usadel equations in useful representations. Now let us use the assumption that the field is monochromatic with a frequency $\omega_0$: $A\left(\omega \right)=2\pi \delta \left(\omega -\omega _{0} \right)A_{+} +2\pi \delta \left(\omega +\omega _{0} \right)A_{-} $. From the periodicity of the field, it follows that Green's functions and other frequency-dependent quantities can be expanded in harmonics of \textit{$\omega $}${}_{0}$:
\[G_{\omega } \left(E\right)=\sum _{n}2\pi \delta \left(\omega -n\omega _{0} \right)G_{n} \left(E\right) \]
The terms with \textit{A} now become
\[\left(A\circ \hat{G}\circ A\circ \hat{G}\right)_{n} \left(E\right)=\sum _{s_{1} ,s_{2} ,m}A_{s_{1} } \hat{G}_{m-s_{1} } \left(E_{-s_{1} -m+n} \right)A_{s_{2} } \hat{G}_{n-m-s_{2} } \left(E_{-s_{2} -m} \right) ,\]
\[\left(\hat{G}\circ A\circ \hat{G}\circ A\right)_{n} \left(E\right)=\sum _{s_{1} ,s_{2} ,m}\hat{G}_{m-s_{1} } \left(E_{+s_{1} -m+n} \right)A_{s_{1} } \hat{G}_{n-m-s_{2} } \left(E_{+s_{2} -m} \right) A_{s_{2} } ,\]
where indexes ${s}_{1,2}$ runs over $\pm$1 and $E_k \equiv E+k{\omega}_{0}/2$.

The other terms in the retarded Usadel equation are evaluated much simpler and their harmonics are
\[\left(\hat{\tau }_{3} \partial _{t_{1} } \hat{G}+\partial _{t_{2} } \hat{G}\hat{\tau }_{3} \right)_{n} \left(E\right)=-iE\left(\hat{\tau }_{3} \hat{G}{}_{n} \left(E\right)-\hat{G}{}_{n} \left(E\right)\hat{\tau }_{3} \right)-i\frac{n\omega _{0} }{2} \left(\hat{\tau }_{3} \hat{G}{}_{n} \left(E\right)+\hat{G}{}_{n} \left(E\right)\hat{\tau }_{3} \right),\]
\[\left(\hat{G}\circ \hat{G}\right)_{n} \left(E\right)=\sum _{m}\hat{G}_{m} \left(E_{-m+n} \right)\hat{G}_{n-m} \left(E_{-m} \right) ,\]
\[\left(\hat{\Delta }\circ \hat{G}\right)_{n} \left(E\right)=\sum _{m}\hat{\Delta }_{m} \hat{G}_{n-m} \left(E_{-m} \right) ,\]
\[\left(\hat{G}\circ \hat{\Delta }\right)_{n} \left(E\right)=\sum _{m}\hat{G}_{n-m} \left(E_{+m} \right)\hat{\Delta }_{m}  .\]

%To have an ability to consider more general and realistic case let us add also a term describing constant depairing. Such a term is required to account for smearing of singularities of Green functions which in reality take place even with no rf field. This depairing can be origined by magnetic impurities or low-scale (of order $\xi$) inhomogenities, or can be a manifestation of limitations of BCS theory. It is usually assumed that the source of depairing acts as a dc supercurrent, thus to account for it phenomenologically, the additional depairing term can be written as
%	\[
%\]
%(Obviously, an "`onest"' example of situation where constant depairing term arising essentialy in this form is a case where besides rf field where exists also dc supercurrent described by vector-potential $A_0$ (to avoid interplay between dc and rf current and arising of cross-terms containing both $A_0$ and $A_{\pm}$ one can apply current perpendicularly to rf field). In this case, $\Gamma=2 e^2 A_0^2 D$).

%Besides greater realisity, introducing of constant depairing term allows to obtain some analytical results, thanks to smearing singularities of Green functions.

Collecting all the terms and restoring matrices and coefficients, one obtains a Usadel equation in the form

\begin{equation} \label{R-Usadel_E-n}
\begin{array}{c}
{e}^{2}D \sum_{s_{1} ,s_{2} ,m} A_{s_{1}}A_{s_{2}}\left(\hat{\tau}_{3}\hat{G}_{m-s_{1} } \left(E_{-s_{1} -m+n} \right) \hat{\tau}_{3}\hat{G}_{n-m-s_{2} } \left(E_{-s_{2} -m}\right) - \hat{G}_{m-s_{1} } \left(E_{+s_{1} -m+n} \right)\hat{\tau}_{3} \hat{G}_{n-m-s_{2} } \left(E_{+s_{2} -m} \right) \hat{\tau}_{3}\right)-
%\\
%+\frac{\Gamma}{2}\sum_{m}\left(\hat{\tau }_{3}\hat{G}_{m} \left(E_{-m+n} \right)\hat{\tau }_{3}\hat{G}_{n-m} \left(E_{-m} \right)-
%\hat{G}_{m} \left(E_{-m+n} \right)\hat{\tau }_{3}\hat{G}_{n-m} \left(E_{-m}\right)\hat{\tau }_{3} \right)-
\\
 -iE\left[\hat{\tau }_{3}, \hat G_n \left(E\right)\right]-i\frac{n\omega _{0}}{2} \left\{\hat{\tau }_{3}, \hat G_n \left(E\right)\right\} - i\sum _{m}\left(\hat{\Delta }_{m} \hat{G}_{n-m} \left(E_{-m} \right)-\hat{G}_{n-m} \left(E_{+m} \right)\hat{\Delta }_{m}\right)=0,
\end{array}
\end{equation}

One sees that the equation for harmonics is a set of linked equations at every $n$. There also exists an equal number of equations emerging from the normalization condition $\hat{G}\circ \hat{G}=\hat{\delta }\left(t_{1} -t_{2} \right)$, whose form in frequency-energy representation is

\begin{equation} \label{NormC_E-n}
\sum _{m}\hat{G}_{m} \left(E_{-m+n} \right)\hat{G}_{n-m} \left(E_{-m} \right) =\hat{\delta }_{n,0} .
\end{equation}

In scalar form, the Usadel equation and normalization condition are

\begin{equation} \label{R-Usadel_E-ns}
\begin{array}{c}
{e}^{2}D \sum_{s_{1} ,s_{2} ,m} A_{s_{1}}A_{s_{2}}
( G_{m-s_{1} } \left(E_{-s_{1} -m+n} \right) G_{n-m-s_{2}}\left(E_{-s_{2} -m}\right)+
F_{m-s_{1} } \left(E_{-s_{1} -m+n} \right) F_{n-m-s_{2}}\left(E_{-s_{2} -m}\right)-
\\
- G_{m-s_{1} } \left(E_{+s_{1} -m+n} \right)G_{n-m-s_{2} } \left(E_{+s_{2} -m} \right)
- F_{m-s_{1} } \left(E_{+s_{1} -m+n} \right)F_{n-m-s_{2} } \left(E_{+s_{2} -m} \right) )-
\\
 -i n\omega _{0} G_n \left(E\right) -
 i\sum _{m}{\Delta }_{m}\left(F_{n-m} \left(E_{-m} \right)-F_{n-m} \left(E_{+m} \right)\right)=0,
\\
{e}^{2}D \sum_{s_{1} ,s_{2} ,m} A_{s_{1}}A_{s_{2}}
( G_{m-s_{1} } \left(E_{-s_{1} -m+n} \right) F_{n-m-s_{2}}\left(E_{-s_{2} -m}\right)+
F_{m-s_{1} } \left(E_{-s_{1} -m+n} \right) G_{n-m-s_{2}}\left(E_{-s_{2} -m}\right)+
\\
+ G_{m-s_{1} } \left(E_{+s_{1} -m+n} \right)F_{n-m-s_{2} } \left(E_{+s_{2} -m} \right)
+ F_{m-s_{1} } \left(E_{+s_{1} -m+n} \right)G_{n-m-s_{2} } \left(E_{+s_{2} -m} \right) )-%+
%\\
%+\Gamma\sum_{m}\left(G_m \left(E_{-m+n} \right)F_{n-m} \left(E_{-m} \right)+
%F_m \left(E_{-m+n} \right)G_{n-m} \left(E_{-m}\right)\right)-
\\
-2 i E F_n \left(E\right)-i\sum _{m}{\Delta }_{m}\left(G_{n-m} \left(E_{-m} \right)+G_{n-m} \left(E_{+m} \right)\right)=0.
\end{array}
\end{equation}

\begin{equation} \label{NormC_E-ns}
\begin{array}{c}
\sum _{m}\left(G_m \left(E_{-m+n} \right)G_{n-m} \left(E_{-m} \right)-F_m \left(E_{-m+n} \right)F_{n-m} \left(E_{-m} \right)\right) =\delta_{n,0} ,
\\
\sum _{m}\left(G_m \left(E_{-m+n} \right)F_{n-m} \left(E_{-m} \right)-F_m \left(E_{-m+n} \right)G_{n-m} \left(E_{-m} \right)\right) = 0 .
\end{array}
\end{equation}

Green's functions at nonzero frequencies emerge only if the field is nonzero and tend to 0 continuously with the decrease of the field, so one can expand $G_{n}$ by powers of $A$. From the ''conservation of frequency'' (frequencies at the left- and right-hand sides of all equalities should be equal), it follows that the main term in the series for $G_{n}$ is at least $A^{n}$. On the other hand, because $A$ is a vector and $G$ a scalar, and there is no other vector except $A$ in the problem, series for $G$ can contain only even powers of $A$. So it is obvious that $G_n$ with odd $n$ vanish. Thus one should keep only terms with even indexes of the Green's functions in the above-written expressions. %\textit{(What to vector quantities like current, directly opposite situation takes place for them because they contain an additional power of $A$ -- they have only odd harmonics.)}

Equations \eqref{R-Usadel_E-n} and \eqref{NormC_E-n} or \eqref{R-Usadel_E-ns} and \eqref{NormC_E-ns}, being supplemented by the self-consistency equation for $\hat\Delta$%(TO BE ADDED)
, form a complete set describing the problem. %As was mentioned before,
The set is redundant: upper and lower equations in \eqref{R-Usadel_E-ns} and \eqref{NormC_E-ns} contain identical information (below this will be demonstrated explicitly for a particular case), expect the case $n=0$ when left-hand sides of upper equation \eqref{R-Usadel_E-ns} and lower equation \eqref{NormC_E-ns} are identically zeros.\\ %I suppose they can be solved numerically with the use of conventional Newton method.

\textbf{\S 2. Closed Usadel equation for time-averaged Green functions}.

To obtain a closed equation for the harmonic at zero frequency, one can consider the amplitude of the vector-potential $A$ as a small parameter and, keeping only terms not greater than $A^{2}$, obtain for $n=0$
\[\left(A\circ \hat{G}\circ A\circ \hat{G}\right)_{0} \left(E\right)\approx A_{+} \hat{G}_{0} \left(E_{-2} \right)A_{-} \hat{G}_{0} \left(E\right)+A_{-} \hat{G}_{0} \left(E_{+2} \right)A_{+} \hat{G}_{0} \left(E\right),\]
\[\left(\hat{G}\circ A\circ \hat{G}\circ A\right)_{0} \left(E\right)\approx \hat{G}_{0} \left(E\right)A_{+} \hat{G}_{0} \left(E_{-2} \right)A_{-} +\hat{G}_{0} \left(E\right)A_{-} \hat{G}_{0} \left(E_{+2} \right)A_{+} ,\]
and the retarded Usadel equation

\begin{equation} \label{R-Usadel_E-0}
-iE\left[\hat{\tau }_{3} ,\hat{G}{}_{0} \right]-i\left[\hat{\Delta }_{0} ,\hat{G}{}_{0} \right]+\alpha \left(\hat{\tau }_{3} \left(\hat{G}{}_{0+} +\hat{G}{}_{0-} \right)\hat{\tau }_{3} \hat{G}{}_{0} -\hat{G}{}_{0} \hat{\tau }_{3} \left(\hat{G}{}_{0+} +\hat{G}{}_{0-} \right)\hat{\tau }_{3} \right)=0,
\end{equation}

\noindent where $\hat{G}_{0} \equiv \hat{G}_{0} \left(E\right)$, $\hat{G}_{0\pm } \equiv \hat{G}_{0} \left(E_{\pm 2} \right)=\hat{G}_{0} \left(E\pm \omega _{0} \right)$ and $\alpha \equiv {e}^{2} D {A}_{+} {A}_{-}$, and the normalization condition

\begin{equation} \label{NormC_E-0}
\hat{G}_{0} ^{2} =\hat{1}.
\end{equation}

One can see that the equations are closed with respect to ${G}_{0}$. This is due to the fact that, for $n=0$, all the terms containing ${G}_{m \neq 0}$, both in the Usadel equation \eqref{R-Usadel_E-n} and in the normalization condition \eqref{NormC_E-n}, are at least of 4-th order in $A$.

In scalar form, the equations can be obtained by substituting
\[\hat{G}_{0} =\left(\begin{array}{cc} {G_{0} } & {F_{0} } \\ {-F_{0} } & {-G_{0} } \end{array}\right)\equiv G_{0} \hat{\tau }_{3} +F_{0} i\hat{\tau }_{2} \]
and are
\begin{equation} \label{R-Usadel_E-0s}
-iEF_{0}-i\Delta G_{0}+\alpha \left\{\left(G_{0+} +G_{0-} \right)F_{0}+\left(F_{0+} +F_{0-} \right)G_{0}\right\}=0
\end{equation}
and
\begin{equation} \label{NormC_E-0s}
G_{0}^{2} -F_{0}^{2} =1.
\end{equation}

These are the equations we work with in the main text.\\

%Here, for comparison, is the retarded Usadel equation in the case of depairing by dc current

%\[
%-iEF_{0}-i\Delta G_{0}+\Gamma G_{0}F_{0}=0
%\]

\textbf{Validity criterion of closed equations for the zero-frequency harmonic.}

To use closed equations for zero-frequency harmonic $\hat G_0$, one should find a criterion of smallness of $A$. A natural way to do this is to find $\hat G_{2}$ keeping only $\propto A^{0}$ and $\propto A^{2}$ terms in \eqref{R-Usadel_E-n} and \eqref{NormC_E-n} and then see at which $A$ the condition $\hat G_{2} \ll \hat G_{0}$ holds. To find $\hat G_{2}$, one should solve the Usadel equation at $n=2$, which with accuracy up to terms $\propto A^{2}$ is:
\begin{equation} \label{R-Usadel_E-2}
%\begin{array}{c}
{\alpha}_{+}\left(\hat{\tau}_{3}\hat{G}_{0}\hat{\tau}_{3}\hat{G}_{0-}-\hat{G}_{0+}\hat{\tau}_{3} \hat{G}_{0}\hat{\tau}_{3}\right)%+
%\frac{\Gamma}{2}\left(\hat{\tau}_{3}\hat{G}_{2}\hat{\tau}_{3}\hat{G}_{0-}+
%\hat{\tau}_{3}\hat{G}_{0+}\hat{\tau}_{3}\hat{G}_{2}-
%\hat{G}_{2}\hat{\tau}_{3} \hat{G}_{0-}\hat{\tau}_{3}
%-\hat{G}_{0+}\hat{\tau}_{3} \hat{G}_{2}\hat{\tau}_{3}\right)-
%\\
 -iE\left[\hat{\tau }_{3}, \hat{G}{}_{2} \right]
 -i\omega _{0} \left\{ \hat{\tau }_{3}, \hat{G}{}_{2}\right\}
 -i\left[\hat{\Delta }_{0}, \hat{G}_{2}\right]
 -i\left(\hat{\Delta }_{2} \hat{G}_{0-}-\hat{G}_{0+}\hat{\Delta}_{2}\right)=0,
%\end{array}
\end{equation}
where $\hat G_0$ is written in zero'th order accuracy in $A$, i.e. it is a solution of the Usadel equation \eqref{R-Usadel_E-0} with $\alpha=0$, and $\alpha_{+}\equiv e^2 D A_{+}^2$ (it can differ from $\alpha$ only by a phase factor). In scalar form, this results in two linear equations with respect to $G_2$, $F_2$ ($\hat{G}_2 = G_2\hat{\tau }_{3} +F_2 i\hat{\tau }_{2}$):
\begin{equation} \label{R-Usadel_E-2s}
\begin{array}{c}
-2i\omega_{0}G_2=
-{\alpha}_{+}\left(G_0\left({G}_{0-}-G_{0+}\right)+F_{0}\left({F}_{0-}-F_{0+}\right)\right)+
 i{\Delta }_{2}\left({F}_{0-}-F_{0+}\right),
\\
 -2iEF_{2}-2i{\Delta}_0 G_{2}%+
% \Gamma \left(\left({G}_{0-}+G_{0+}\right)F_{2}+\left({F}_{0-}+F_{0+}\right)G_{2}\right)=
%\\
 =-{\alpha}_{+}\left(G_0\left({F}_{0-}+F_{0+}\right)+F_{0}\left({G}_{0-}+G_{0+}\right)\right)+
 i{\Delta }_{2}\left({G}_{0-}+G_{0+}\right).
\end{array}
\end{equation}
The first equation is the component of matrix equation \eqref{R-Usadel_E-2} proportional to $\hat 1$ and the second one is that proportional to $\hat{\tau }_{3}$. (Two other components are identically zero).
The normalization condition for $n=2$ with terms up to $\propto A^{2}$ reads
\[\hat{G}_{2}\hat{G}_{0-}+\hat{G}_{0+}\hat{G}_{2}=0,\]
(largest neglected terms are $\propto A^{6}$)
or in a scalar form
\begin{equation} \label{NormC_E-2s}
\begin{array}{c}
\left(G_{0-}+G_{0+}\right)G_{2}-\left(F_{0-}+F_{0+}\right)F_{2}=0,
\\
\left(F_{0-}-F_{0+}\right)G_{2}-\left(G_{0-}-G_{0+}\right)F_{2}=0.
\end{array}
\end{equation}
The determinant of the system is zero as one can prove using the normalization condition for $n=0$ \eqref{NormC_E-0s}, so the equations are identical. The two equations \eqref{R-Usadel_E-2s} are also identical: dividing the second of them on the first, using normalization condition \eqref{NormC_E-2s}, and rearranging terms, one obtains the equality
%\[
%\left(E+i\frac{\Gamma}{2} \left({G}_{0-}+G_{0+}\right)\right)\left(F_{0-}-F_{0+}\right)+
%\left(\Delta_0+i\frac{\Gamma}{2} \left({F}_{0-}+F_{0+}\right)\right)\left(G_{0-}+G_{0+}\right)-
%\omega_0\left(F_{0-}+F_{0+}\right)=0
%\]
\[
E\left(F_{0-}-F_{0+}\right)+
\Delta_0\left(G_{0-}+G_{0+}\right)-
\omega_0\left(F_{0-}+F_{0+}\right)=0
\]
which is nothing except the difference of two Usadel equation for zero frequency (in zero'th order in $A$), written for energies $E \pm \omega_0$. Thus one can choose, for instance, firsts of the equations \eqref{R-Usadel_E-2s} and \eqref{NormC_E-2s} and solve the obtained system to find $G_2$ and $F_2$. The solution is
\begin{equation} \label{G_2}
\begin{array}{c}
G_2=
\frac{-{\alpha}_{+}\left(G_0\left({G}_{0-}-G_{0+}\right)+F_{0}\left({F}_{0-}-F_{0+}\right)\right)+
 i{\Delta }_{2}\left({F}_{0-}-F_{0+}\right)}{-2i\omega_{0}},
\\
F_{2}=\frac{G_{0-}+G_{0+}}{F_{0-}+F_{0+}}G_2 .
\end{array}
\end{equation}

To find when the condition of smallness of $G_2$ is fulfilled, note that
differences $G_{0-}-G_{0+}$, $F_{0-}-F_{0+}$ are of order unity and to have $G_2, F_2 \ll G_0, F_0$, one requires
	\[\alpha \ll \omega_0.\] \\
%\textit{(THIS IS TOO ROUGH. INDEED, $G_0$, $F_0$ are of order $\alpha^{1/2}(\omega_0 \Delta)^{-1/4}$ at the point $E=\Delta$ (analytics shows it), thus for $E=\pm\Delta \pm \omega_0$ differences $G_{0-}-G_{0+}$, $F_{0-}-F_{0+}$ are large compared to unity. But resulting condition of smallness of $G_2$, $F_2$ remains the same.)}
%In the opposite case $\omega_0 \ll \delta \epsilon$, $G_{0-}-G_{0+}, F_{0-}-F_{0+} \leq \omega_0/\delta \epsilon$ and the condition is more soft,
%%	\[\alpha \ll \delta \epsilon.
%\]
 %One can see that
%(FORMULA EXPRESSING $\delta \epsilon$ VIA $\Gamma$ and $\Delta$ WILL BE ADDED).

\textbf{\S 3. Expressions for some observable quantities.}

Zero'th order harmonics Green's functions $G_0$, $F_0$ have the sense of time-averaged Green's functions. Time-averaged values of spectral quantities which are linear in the Green's functions, can be calculated simply by replacing $G$, $F$ to $G_0$, $F_0$ in corresponding formulas. For instance, the time-averaged density of states is $N(E)=N_0\Re G_0$. %Fig. 1 presents calculated DOS for couple of values of $\alpha$ and $\omega$. For comparison, at the same plot there also presented DOS for the cases of zero field (standard BSC DOS) and depairing by dc current.

For the (linear) conductivity at frequency $\omega$, one obtains

\begin{equation} \label{sigmaS}
\sigma(\omega)=\frac{\sigma_N}{4\omega}\int d E \left\{\left(G_{0-}^{*}\Re G_0+F_{0-}^{*}\Re F_0\right)f_L+
\left(G_0\Re G_{0-}+F_0\Re F_{0-}\right)f_{L-}\right\},
\end{equation}

where $\sigma_N=2 e^2 N_0 D$.
For $\omega \ll \Delta$, formula for imaginary part of conductivity reduces to

\begin{equation} \label{imsigma}
\Im\sigma(\omega)=\frac{\sigma_N}{2\omega}\int d E \left(\Re G_0\Im G_0+\Re F_0\Im F_0\right)f_L.
\end{equation}

\textbf{Numerical solution of closed equations for zero-frequency harmonic.}

To solve numerically the equation for the time-averaged Green's functions (\ref{R-Usadel_E-0s}) together with normalizing condition (\ref{NormC_E-0s})  we used a direct iterative procedure.
We rewrite Eq. (\ref{R-Usadel_E-0s}) as

\begin{equation} \label{num1}
-i\tilde{E}F_{0}-i\tilde{\Delta} G_{0}=0
\end{equation}

with

\begin{equation} \label{num2}
\tilde{E}=E+i\alpha \left(G_{0+} +G_{0-} \right),~\tilde{\Delta}=\Delta+i\alpha \left(F_{0+} +F_{0-} \right).
\end{equation}

We substitute $G_0$ and $F_0$ found at the previous iteration into the formulas for $\tilde{E}$ and $\tilde{\Delta}$ (\ref{num2}) and then solve Eq. (\ref{num1}) to find the new $G_0$ and $F_0$.
Taking into account the normalization condition (\ref{NormC_E-0s}) the solution of Eq. (\ref{num1}) has a BCS-like form: $G_0=\tilde{E}/\sqrt{\tilde{E}^2-\tilde{\Delta}^2},~F_0=-\tilde{\Delta}/\sqrt{\tilde{E}^2-\tilde{\Delta}^2 }$.
At the first iteration, when BCS Green's functions are used, we calculate $\tilde{E}$  and $\tilde{\Delta}$  using (\ref{num2}) only for energies $|E \pm \Delta| \leq a\hbar\omega_0$  ($a$ is the number smaller than unity, we took $a=0.75$) and set $\tilde{E}=E,~\tilde{\Delta}=\Delta$,   outside this interval to avoid BSC singularities. At the next iterations the behavior of the Green's functions became smooth near $\Delta$, so we simply iterate Eqs. (\ref{num1}) and (\ref{num2}) until we come to appropriate accuracy. 20 cycles were enough to reach convergence.

The self-consistent  $\Delta$ in the presence of rf field remains finite. To generate plots for the DOS we did not use directly the self-consistency equation for the order parameter (Eq. 5 of the main text) and neglected the deviation of $\Delta$ compared to its value without rf field $\Delta_0$. At the values of $\alpha$ used for this calculation the change of $\Delta$ is too small to be seen on those DOS plots. However, the (self-consistent) change of  $\Delta$ was included in the calculation of the correction to the conductivity, in particular while deriving the factor in Eq. 7. Because we are interested in a small correction linear in $\alpha$, we exploited the expansion for small correction  $\delta \Delta$:

\begin{equation} \label{num3}
\delta \Delta=\alpha \left( -\frac{\lambda}{2} \int dE \frac{d \Re {F}}{d\alpha}f_L\right)=\left( -\frac{\lambda}{2} \int dE \frac{  \partial \Re {F}}{\partial \alpha}_{|\Delta=\Delta_0} f_L\right)\alpha+\left( -\frac{\lambda}{2} \int dE \frac{  \partial \Re {F}}{\partial \Delta}_{|\alpha=0} f_L\right)\delta\Delta.
\end{equation}

After evaluation of the integral in the second term, this yields $\delta\Delta=\left( -\frac{1}{2} \int dE \frac{  \partial \Re {F}}{\partial \alpha}_{|\Delta=\Delta_0} f_L\right)\alpha$.

To calculate the correction to the kinetic inductance at low frequency $\omega \ll \Delta$, we present it as

\begin{equation} \label{num4}
\frac{\delta L_k}{L_k}=-\frac{\delta \Im{\sigma}}{\Im{\sigma}}=-\frac{1}{\Im{\sigma}}\left( \frac{  \partial \Im {\sigma}}{\partial \alpha}_{|\Delta=\Delta_0} \alpha + \frac{  \partial \Im {\sigma}}{\partial \Delta}_{|\alpha=0} \delta\Delta \right)=\frac{1}{\Im{\sigma}}\frac{  \partial \Im {\sigma}}{\partial \alpha}_{|\Delta=\Delta_0}\alpha-\frac{\delta\Delta}{\Delta}.
\end{equation}

The second term in (\ref{num4}) is the change of the kinetic inductance of the superconductor, caused by the change of $\Delta$, without rf field. The first term in (\ref{num4}) is evaluated using the formula for the imaginary part of the conductivity (\ref{imsigma})

\begin{equation} \label{num5}
\frac{  \partial \Im {\sigma}}{\partial \alpha}_{|\Delta=\Delta_0}\alpha=\left(\int dE \frac{  \partial \Im {F^2}}{\partial \alpha} f_L\right)\alpha.
\end{equation}

The integral in (\ref{num5}) and the integral in the expression for $\delta\Delta$ (\ref{num3}) are evaluated numerically, with the replacement of derivatives by $\alpha$ to finite differences. The distribution function $f_L$ was set to $sign(E)$, which corresponds to the low-temperature limit. As a result for $\delta L_k/L_k$, we obtain the formula (7) of the main text, i.e. calculate the numerical factor between $\delta L_k/L_k$ and $\alpha/\Delta$.\\

\textbf{Relation to Floquet states}

Although we do not work in the basis of Floquet states, the Hamiltonian with a time-dependent periodic term and especially the resulting coherent excited states are clearly reminiscent of Floquet states \cite{grif}.

%\begin{thebibliography}{99}

%\bibitem {lo} A. I. Larkin, Yu. N. Ovchinnikov, ZhETP \textbf{73}, 299 (1977).

%\bibitem{grif}M. Grifoni and P.H\"anggi, Physics Reports \textbf{304}, 229
(1998).

%\end{thebibliography}

\end{document}